\documentclass{elsarticle}
\usepackage{latexsym}
\usepackage{float}
\usepackage[latin2]{inputenc}
\usepackage{graphicx}
\usepackage{ae}
\usepackage{aecompl}
\usepackage{setspace}
\usepackage{amsmath}
\usepackage{amssymb}
\usepackage{amsthm}
\newtheorem{theorem}{Theorem}

\newtheorem{remark}[theorem]{Remark}

\usepackage[hidelinks]{hyperref}

\makeatletter
\def\ps@pprintTitle{%
	\let\@oddhead\@empty
	\let\@evenhead\@empty
	\def\@oddfoot{\centerline{\thepage}}%
	\let\@evenfoot\@oddfoot}
\makeatother

\begin{document}

\title{The Frequent Subgraphs of the Connectome of the Human Brain}
	
\author[p]{Máté Fellner}
\ead{fellner@pitgroup.org}
\author[p]{Bálint Varga}
\ead{balorkany@pitgroup.org}
\author[p,u]{Vince Grolmusz\corref{cor1}}
\ead{grolmusz@pitgroup.org}
\cortext[cor1]{Corresponding author}
\address[p]{PIT Bioinformatics Group, Eötvös University, H-1117 Budapest, Hungary}
\address[u]{Uratim Ltd., H-1118 Budapest, Hungary}

\date{}

\begin{abstract}
	In mapping the human structural connectome, we are in a very fortunate situation: one can compute and compare graphs, describing the cerebral connections between the very same, anatomically identified small regions of the gray matter among hundreds of human subjects. The comparison of these graphs has led to numerous recent results, as the (i) discovery that women's connectomes have deeper and richer connectivity-related graph parameters like those of men, or (ii) the description of more and less conservatively connected lobes and cerebral regions, and (iii) the discovery of the phenomenon of the Consensus Connectome Dynamics. 	
	Today one of the greatest challenges of brain science is the description and modeling of the circuitry of the human brain. For this goal, we need to identify sub-circuits that are present in almost all human subjects and those, which are much less frequent: the former sub-circuits most probably have functions with general importance, the latter sub-circuits are probably related to the individual variability of the brain structure and functions. The present contribution describes the frequent connected subgraphs (instead of sub-circuits) of at most 6 edges in the human brain. We analyze these frequent graphs and also examine sex differences in these graphs: we demonstrate numerous connected sub-graphs that are more frequent in female or the male connectome. While our results describe subgraphs, instead of sub-circuits, we need to note that all macroscopic sub-circuits correspond to an underlying connected subgraph.	
		Our data source is the public release of the Human Connectome Project, and we are applying the data of 426 human subjects in this study.	
\end{abstract}

\maketitle
	
\section*{Introduction} 

\subsection*{Continuous vs. Discrete Approaches:}

Discrete structures in mathematics are non-continuous objects (like bits, integers, graphs, sequences, logic formulae), which were studied and applied extensively in the last one hundred years in numerous contexts with enormous success (e.g., digital data processing and computing). Before the era of discrete mathematics, the mainstream research area of mathematics was the study of continuous mathematical objects (like geometrical objects, or the real and complex numbers and functions). In the study of living organisms, the discovery of discrete structures like the DNA and RNA sequences or the residues forming the poly-peptide chains of proteins has transformed the once descriptive biology into the fastest developing quantitative discipline of science. 

In brain science the anatomical studies or the electro-physiological methods belong to the continuous approaches, using mathematical tools from geometry and classical analysis of functions. With the dawn of magnetic resonance imaging methods in general and the diffusion-weighted imaging in particular, brain science started applying a rich and well-developed area of discrete mathematics: the theory of graphs in the connectomics studies (e.g., \cite{Hagmann2012,Agosta2014,Ball2014,Szalkai2016a}). 

We believe that the possibilities of studying a very complex discrete structure, the human connectome, or braingraph, will give such breakthrough results to brain science as the studies of the discrete structures like the gene sequences and primary protein structures to biology and medicine. For exploiting these possibilities, one needs graph-theoretical insights into the structure of the human connectome.

In the present contribution, we are mapping the most frequent, connected subgraphs of the human braingraph with at most six edges. We also compare frequent subgraphs in male and female connectomes and find numerous significant differences between the frequencies of some of these graphs between the sexes.

The motivation of this study is the possible discovery of some, still unknown, functional brain circuits on the macroscopic scale. Any individual macroscopic brain circuit needs to be a connected graph; therefore, frequently appearing macroscopic circuits correspond to frequently appearing connected subgraphs. However, the connected braingraphs clearly do not necessarily correspond to functional macroscopic brain circuits. Therefore, our approach can discover the underlying graph structure and not the functional circuitry. 

\subsection*{Previous Work}

The data source of our work is the public data release of the Human Connectome Project \cite{McNab2013}. We have computed numerous directed and undirected braingraphs from the diffusion weighted MRI data of the Human Connectome Project, and we have made these data publicly available at the site \url{https://braingraph.org} \cite{Kerepesi2016b, Szalkai2016d,Szalkai2016,Kerepesi2016,Szalkai2015a}. We have compared deep graph theoretic parameters of the lobes and smaller cerebral areas in \cite{Szalkai2017c}. In the works \cite{Szalkai2015,Szalkai2016a} we have shown that women's connectomes have significantly better, deep connectivity-related graph-theoretical parameters than those of men. In \cite{Szalkai2015c} we have proved that the significant advantage in connectivity-related parameters in women's connectomes is due to sex-, and not to the size-differences. 

In the articles \cite{Szalkai2015a,Szalkai2016} we have described the Budapest Reference Connectome Server \url{https://connectome.pitgroup.org}, which generates parameterized consensus connectomes from hundreds of braingraphs. The Budapest Reference Connectome Server led us to the discovery of the phenomenon of the Consensus Connectome Dynamics \cite{Kerepesi2015b}, and consequently, to a novel method for directing the edges of the human connectomes, gained from diffusion weighted MR imaging data \cite{Szalkai2016d,Kerepesi2016,Szalkai2016e}. 

In the present study, we apply 426 braingraphs from 426 human subjects, each on 1015 vertices from \url{https://braingraph.org}. The workflow of constructing these graphs from the Human Connectome Project data is described in detail in \cite{Kerepesi2016b}. 

Our present contribution is related to the works of \cite{Szalkai2015a,Szalkai2016}, mapping the $k$-frequent edges of a set of human connectomes in the Budapest Reference Connectome Server \url{https://connectome.pitgroup.org}, and also to the work of \cite{Kerepesi2015a}, in which we have mapped the individual variability of the connections within brain regions (lobes and smaller areas). In the present work, however, after deleting the graph edges with fewer, than four defining fibers, we are focusing to smaller {\em connected} subgraphs, which are present in at least the 90\% of the connectomes examined. Therefore, we are describing small, connected, frequent subgraphs with strongly defined edges (i.e., with at least four defining fibers) of the human brain, and we believe that some of these subgraphs also describe relevant functional circuits of the brain.

\section*{Discussion and Results}

The edges of braingraphs connect small gray matter areas of the brain, frequently called ``Regions of Interests'', ROIs. Two ROIs are connected by an edge if axonal fiber tracts are discovered running between them in the tractography phase of the processing workflow. We call these fiber tracts, or, in short, fibers, {\em defining fibers} for the edge in question.

Gray matter can be found on the outer surface of the brain (the cerebral cortex), and also in sub-cortical structures, such as the thalamus, hypothalamus, basal ganglia. The sub-cortical gray matter structures are strongly connected to each other and also to the cortical areas. Therefore, it is not surprising that the most frequent graphs contain predominantly sub-cortical gray matter areas as vertices: e.g., the graph on Figure 1, or the length-2 path \{Left-Caudate,Left-Thalamus-Proper\}\{Left-Thalamus-Proper,Left-Putamen\} have larger than 90\% of frequency of appearance in the connectomes examined.

Consequently, the frequent subgraphs with not only sub-cortical nodes are of special interest, e.g., the edges \{rh.precentral\_7,Right-Putamen\} or
\{rh.caudalmiddlefrontal\_11,Right-Putamen\}, and larger, connected frequent subgraphs, like the star \{Right-Caudate,Left-Caudate\}\{Left-Caudate,Left-Putamen\}\{lh.caudalanteriorcingulate\_2,Left-Caudate\}, or the \{Left-Thalamus-Proper,Left-Caudate\}\{lh.rostralanteriorcingulate\_2,Left-Caudate\}\{Left-Thalamus-Proper,Brain-Stem\},
all with at least 90\% support.

It is an intriguing problem to clarify the roles of the relatively few, specific cortical areas (e.g., lh.rostralanteriorcingulate\_2 as a part of the anterior cingulate cortex), which are connected to the sub-cortical structures in these very frequent connected subgraphs. 

Supporting Table S1 contains all the connected subgraphs of 6 edges or less with frequency of at least 80\%, supporting Table S2 contains those with frequency at least 90\%.

 \begin{table}
 	\centering
 	\begin{tabular}{ l r r r r r r }
 		& $1$ & $2$ & $3$ &$4$ & $5$ & $6$ \\
 		\hline
 		0.9 (all)   &    25 & 71  & 197  & 488  & 1085  & 2117 \\
 		0.8 (all)   &    49 & 179 & 642  & 2168 & 6570  & 17866\\
 		0.7 (all)   &    75 & 324 & 1373 & 5859 & 22774 & 79580\\
 		0.9 (female)&    26 & 73  & 217  & 561  & 1272  & 2529\\
 		0.8 (female)&    49 & 187 & 707  & 2532 & 7937  & 22080\\
 		0.7 (female)&    74 & 341 & 1528 & 6788 & 27560 & 100656\\
 		0.9 (male)  &    26 & 75  & 217  & 542  & 1200  & 2397\\
 		0.8 (male)  &    47 & 178 & 658  & 2320 & 7270  & 20327\\
 		0.7 (male)  &    79 & 354 & 1464 & 6194 & 23977 & 84934
 	\end{tabular}
\caption{The number of the $k$-element connected subgraphs with different minimum support and input set choices. Values for $k=1,2,...,6$ are given in the columns, while the rows correspond to the minimum supports of 0.9, 0.8 and 0.7, and the input sets with all, only the female, and only the male subjects. It is interesting to note that the number of frequent 1-sets are about the same for all three input sets, the number of frequent length-2 paths are larger for males for 70\% and 90\% frequencies, while the number of frequent, connected 6-sets are much larger in the case of female connectomes, for all three frequency bounds.}
\end{table}

\begin{figure}[H]
	\begin{center}
		\includegraphics[width=12cm]{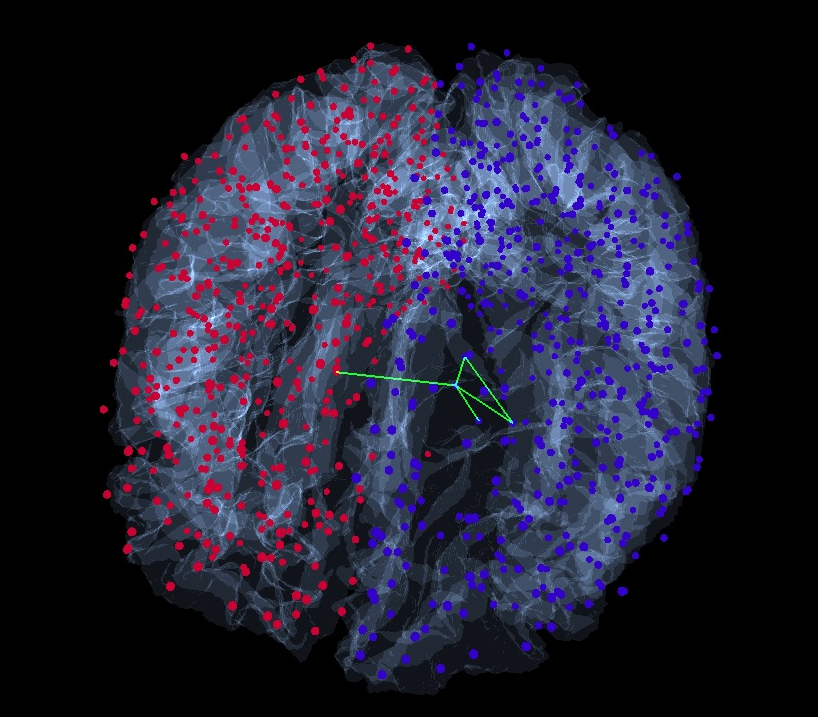}
		\caption{A connected subgraph with five edges, which is present in more than 90\% of human connectomes. Nodes in the left hemisphere are red, in the right hemisphere are blue. The edges of the depicted subgraph are (from left to right): \{Left-Caudate, Right-Caudate\}, \{Right-Caudate, Right-Pallidum\}, \{Right-Caudate, Right-Putamen\},  \{Right-Putamen, Right-Thalamus-Proper\}, \{Right-Thalamus-Proper, Right-Caudate\}. The supporting Table S1 contains all the connected subgraphs with 6 edgfes or less with frequency of at least 80\%, supporting Table S2 contains those with frequency at least 90\%.}
\end{center}
\end{figure}

In our previous works, we have compared numerous deep, graph theoretical parameters of the connectomes of the sexes, and we have found that the female connectome shows much better connectivity-related parameters than those of males \cite{Szalkai2015,Szalkai2016a,Szalkai2015c}. 

In the present work, we have compared the frequencies of the connected subgraphs of at most 6 edges in the male and the female braingraphs. Table 1 lists the number of frequent subgraphs of at most 6 edges for male and female connectomes. It is important to note that the number of frequent 1-sets are about the same for all three input sets, the number of frequent length-2 paths are larger for males for 70\% and 90\% frequencies, while the number of frequent, connected 6-sets are much larger in the case of female connectomes, for all three frequency bounds. This observation independently strengthens our results in \cite{Szalkai2015,Szalkai2016a,Szalkai2015c}, showing that women have ``better connected'' braingraphs than men.

The supporting Table S3 contains the description of the subgraphs with different frequencies, and also the significance of the frequency differences. Figure 2 shows an example for a frequent connected subgraph, which is significantly more frequent in female than in male connectomes. Figure 3 shows an example for a connected subgraph that is significantly more frequent in male braingraphs than in female ones.

\begin{figure}[H]
	\begin{center}
		\includegraphics[width=12cm]{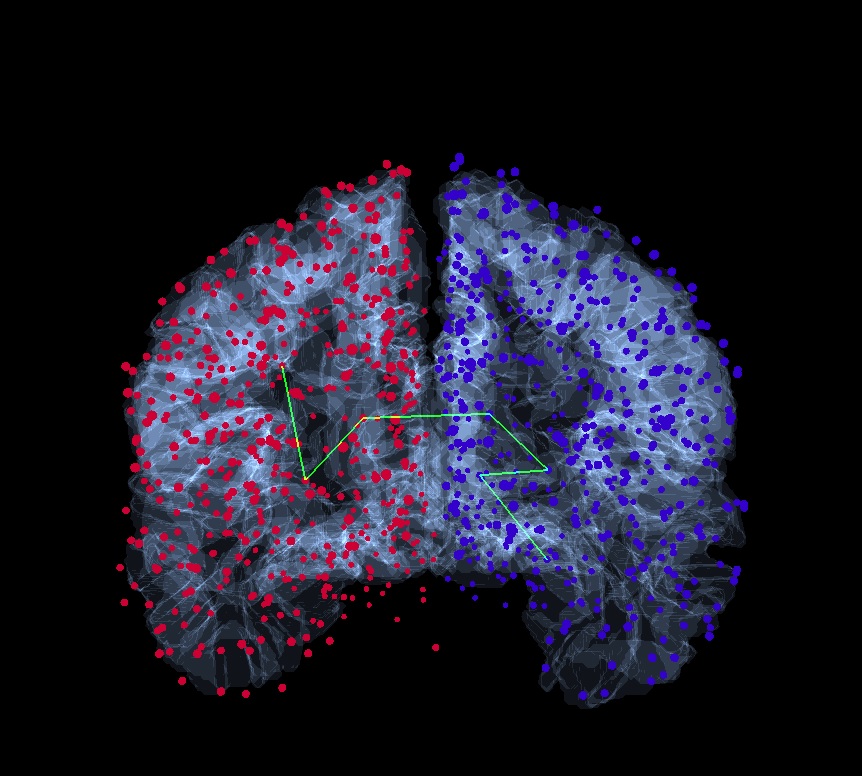}
		\caption{A connected subgraph with six edges, which is more frequent in female than in male braingraphs (male frequency: 74.03\%, female frequency: 90.08\%, $p=10^{-5}$). Nodes in the left hemisphere are red, in the right hemisphere are blue. The edges of the depicted subgraph are (from left to right): \{lh\_precentral\_6, Left-Putamen\}, \{ Left-Putamen, Left-Caudate\}, \{Left-Caudate, Right-Caudate\},  \{Right-Caudate, Right-Putamen\}, \{Right-Putamen, Right-Thalamus-Proper\}, \{Right-Thalamus-Proper, Right-Hippocampus\}. Supporting Table S3 contains those connected subgraphs, which have a minimum support of 90\% in either the male or the female subset, and whose  the supporting Table S3, with the frequencies for male and female braingraphs, and the uncorrected and Holm-Bonferroni corrected p values. }
			
	\end{center}
\end{figure}

\begin{figure}[H]
	\begin{center}
		\includegraphics[width=12cm]{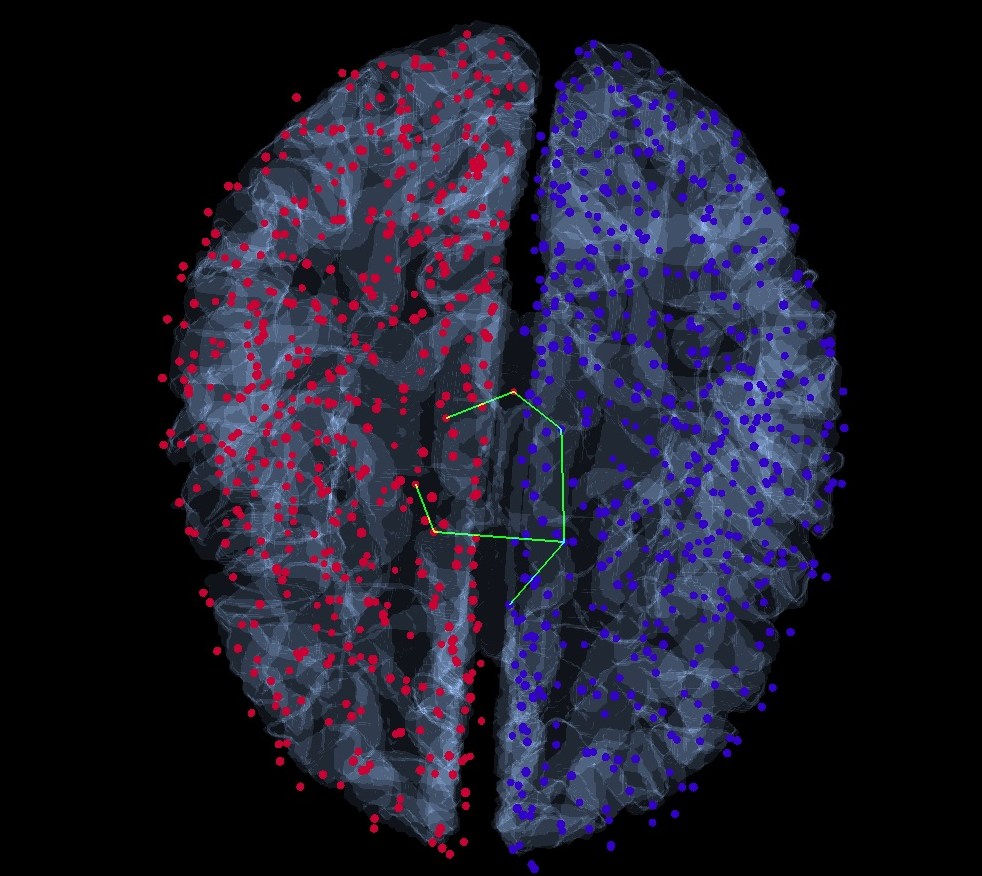}
		\caption{A connected subgraph with six edges, which is more frequent in male than in female braingraphs (male frequency: 91.71\%, female frequency: 83.88\%, $p=0.016$). Nodes in the left hemisphere are red, in the right hemisphere are blue. The edges of the depicted subgraph are listed in clockwise direction: \{Left-Thalamus-Proper, Brain-Stem\}, \{ Brain-Stem, Right-Thalamus-Proper\}, \{Right-Thalamus-Proper, Right-Caudate\},  \{Right-Caudate, rh\_caudalanteriorcingulate\_1\}, \{Right-Caudate, Left-Caudate\}, \{Left-Caudate, Left-Pallidum\}.}
		
	\end{center}
\end{figure}

\section*{Materials and Methods}

The dataset consisted of graphML files, (denoted by ``Full set, 426 brains'' was downloaded from the publicly available  \url{braingraph.org} repository of ours, from the \url{https://braingraph.org/download-pit-group-connectomes/} directory. The graphs in the directory were constructed from the Human Connectome Project \cite{McNab2013} public release, as described in detail in \cite{Kerepesi2016b}. For processing the graphML files the NetworkX library was used (available on \url{https://networkx.github.io/} ) in Python (version: 3.6.2.).

For the nomenclature of ROIs, we apply the names used by FreeSurfer \cite{Fischl2012}; the abbreviation rh means ``right hemisphere'', lh means ``left hemisphere''.

In the graph construction, described in \cite{Kerepesi2016b}, a graph edge was added to the graph if at least one fiber tract was discovered in the tractography step of the MRI data processing between its endpoints. 

However, when only a small number of independently discovered fibers define a graph edge, then this edge could be an artifact, so, in this work, we preprocessed the graphs by eliminating graph edges defined by less than 4 fiber tracts. Therefore, in this work, we are considering only ``strong'' graph edges, defined by at least 4 fiber tracts. If an edge had lower fiber number than this value, we considered it a ``weak edge'' and we removed it from the edge list.

We note that this filtering process has consequences for the edge frequency counting. For example, suppose that an edge has 4 fibers in 90\% and 0 fibers in 10\% of the graphs, but another edge has 3 fibers in 21\% and 10 fibers in 79\% of the connectomes, then the former will be considered  in 80\% support even though the latter has more fiber in average, so this filtering process may distort the sample. 

As an example, let us mention the frequent \{rh.insula\_7,Right-Putamen\} edge. It appears in 402 braingraphs, but in only 253 braingraphs contains this edge with at least 4 defining fibers; therefore, the \{rh.insula\_7,Right-Putamen\} edge does not appear in our result sets of either 80\% or 90\% support (i.e., in supporting Tables S1 and S2, respectively).

Additionally, we have also removed the loop edges from the graphs.

In this work, we have applied the famous apriori algorithm \cite{Agrawal1994a} from data mining \cite{han-kamber} for finding the frequent edge sets of the connectomes. 

\subsection*{The apriori algorithm}

The apriori algorithm is a tool in data mining to find frequent items sets and association rules in datasets. Let $\mathcal{I}$ be the set of items, $\mathcal{D}$ be the set of transactions, which are subsets of $\mathcal{I}$, furthermore let there be an $s$ support percentage, where $s \in (0,100]$. The problem is to find every $X \subset \mathcal{I}$ which is a subset of at least $s\%$ of the transactions.

The description of the algorithm is the following:
For every $k \geq 1$ there are two phases, the first phase creates $C_k$, the list of candidate sets with $k$ elements, the second phase counts their support to create $L_k$, the list of candidate sets which are frequent.

First, let $C_1$ be the set of items, so $C_1 = \mathcal{I}$.
For a given $C_k$, and for every $c \in C_k$ the algorithm counts the occurrences of $c$ in the elements of $\mathcal{D}$. So if a $c$ candidate set is contained in $b$ different transactions, then its support will be $b/d$  where $d$ is the size of $\mathcal{D}$. If this $b/d \geq s\%$, then it appends $c$ to $L_k$.
The other step is to create $C_{k+1}$ from $L_k$, for every $l_1$, $l_2$ pair of elements from $L_k$, if the first $k-1$ element of these are the same, then it appends the union of $l_1$ and $l_2$ to $C_{k+1}$ if it has not contained that yet.

The steps can be continued as long as the last $L_k$ is not empty, but in this work, we searched for frequent subsets with at most six elements.

In our application, the itemset $(\mathcal(I))$ is the union of all edges in all the connectomes, and a transaction is a set of edges in a connectome. In our present application, we have 426 connectomes, i.e., 426 transactions.

The code of the algorithm is adapted from the website:  \url{http://adataanalyst.com/machine-learning/apriori-algorithm-python-3-0/}, with changes described below.

\subsection*{Mining frequent connected edge sets}

For finding frequent {\em connected} subgraphs in the set of 426 braingraphs we needed to modify the original apriori algorithm \cite{Agrawal1994a}. First, we need a 

\begin{remark}
	
	If $E_1$ and $E_2$ are connected edge sets which size are $k > 1$ and these sets are equal except their last elements, then $E_1 \cup E_2$ is also a connected edge set.
	
\end{remark}

\begin{proof}
	
	Since $k > 1$, the sets have a common element because by removing their last elements, we get two non-empty, equal sets.
	Let $a_1$ is a node of $E_1$ and $a_2$ is a node of $E_2$ and $b$ is a node of one of the common edges.
	$E_1$ is connected, so there exists a path from $a_1$ to $b$.
	$E_2$ is also connected, so there exists a path from $b$ to $a_2$.
	The union of these paths is a walk, so between every two points of $E_1 \cup E_2$ there is a walk, so the union is connected.
	
\end{proof}

The connected frequent 1-sets are the frequent edges in the connectomes.
The connected 2-sets are the length-2 paths, so for every potential 2-sets, we have to check whether they have a common node. Now, if we have the list of frequent connected $k$-sets, then every frequent $k+1$-set, generated from these, will be connected.

However, we need to make the following

\begin{figure}[H]
	\begin{center}
		\includegraphics[width=12cm]{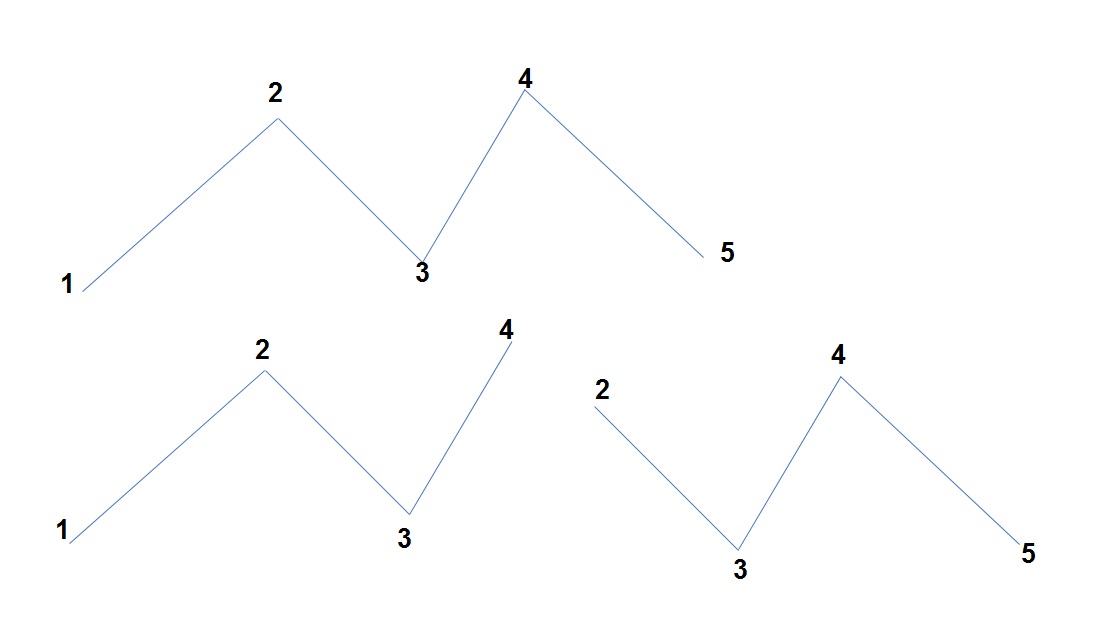}
		\caption{An example, where the original apriori algorithm fails to generate a connected $k+1$-set from two connected $k$-sets.}\label{pelda}		
	\end{center}
\end{figure}

\begin{remark} In the basic apriori algorithm, every $k+1$-set is generated from two specific $k$-set, but there exist connected $k+1$-sets (i.e., connected subgraphs with $k+1$ edges), with one of its generator $k$-set is not connected. For example, (c.f. Fig. \ref{pelda}) consider the following graph: $\{ (1,2), (2,3), (3,4), (4,5) \}$.
As we have seen previously in the general case, the two generators are $\{(1,2),(2,3),(3,4)\}$ and $\{(1,2),(2,3),(4,5)\}$, and the latter is not connected.
\end{remark}

To solve this problem, we have modified the generation step: whenever the union of two $k$-sets from $L_k$ has $k+1$ elements, then we add the union to $C_{k+1}$, if we had not added it before.
In this way, we will not lose any $k+1$-set, because if $E$ is a frequent connected $k+1$ set, then it has at least two connected $k$-subsets which are also frequent, and any two of these clearly generate $E$.
In our example on Fig. \ref{pelda}, these sets are $\{(1,2),(2,3),(3,4)\}$ and $\{(2,3),(3,4),(4,5)\}$.

The running time will increase because previously there were just two generators for a $k+1$ set, now there are $\frac{k(k+1)}{2}$ and every time we have to check the $C_{k+1}$ whether it already contains the set.
However, we believe that this approach is still faster than to perform a breadth-first-search for every frequent edge set.

In our work we searched for small frequent sets, so the maximum $k$ computed was $k = 6$. Table 1 gives the number of the frequent, connected $k$-vertex subgraphs for $k=1,2,...,6$.

The frequent connected subgraphs with 80\% and 90\% support are given in supporting Tables S1 and S2, respectively.

\subsection*{Sex differences}

For discovering differences in the frequencies of connected subgraphs between the female and male connectomes, we have separated the dataset into two parts, to male and female braingraphs. Next we repeated the modified apriori algorithm for both sets.

The comparison of the numbers of frequent connected $k$-sets, found in male and female braingraphs, can be seen in Table 1 for different supports and for $k=1,..., 6$.

Next, we prepared a list of frequent connected graphs which have at least 90\% support in the female or the male sample. 

We have found lots of subgraphs that are more frequent in female connectomes, and several ones that are more frequent in male connectomes. For the determination of the statistical significance of the frequency-differences, we have made a statistical null hypothesis that their frequencies do not differ, and applied $\chi^2$ tests to identify significant differences.

The secondary statistical errors were corrected by the Holm-Bonferroni method \cite{Holm1979}. The results are listed in the supporting Table S3, with the frequencies for male and female braingraphs, and the uncorrected and Holm-Bonferroni corrected p values.

\section*{Conclusions}

We have computed the frequent connected subgraphs of at most 6 edges from 426 human connectomes, applying the data of the Human Connectome Project. We have found that sub-cortical gray matter areas are connected very frequently (at least 90\%) in these graphs, and more interestingly, there are a small number of cortical areas, which are also present in these graphs. We have analyzed sexual differences in the frequency of appearances of these graphs, and have found numerous graphs with significantly higher frequency in female graphs than in male graphs, and a much fewer graphs with significantly higher frequency in male graphs than in female graphs. 

\section*{Data availability} The data source of this study is Human Connectome Project's website at \url{http://www.humanconnectome.org/documentation/S500} \cite{McNab2013}. The connectomes, computed by us can be freely downloaded from   \url{http://braingraph.org/download-pit-group-connectomes/} \cite{Kerepesi2016b}. The large supporting tables can be downloaded from \url{http://uratim.com/freq/Frequent-Supporting.zip}.

\section*{Acknowledgments}
Data were provided in part by the Human Connectome Project, WU-Minn Consortium (Principal Investigators: David Van Essen and Kamil Ugurbil; 1U54MH091657) funded by the 16 NIH Institutes and Centers that support the NIH Blueprint for Neuroscience Research; and by the McDonnell Center for Systems Neuroscience at Washington University. VG and BV were partially supported by the VEKOP-2.3.2-16-2017-00014 program, supported by the European Union and the State of Hungary, co-financed by the European Regional Development Fund, and the NKFI-126472 grant of the National Research, Development and Innovation Office of Hungary.
\bigskip 

\noindent Conflict of Interest: The authors declare no conflicts of interest.



\end{document}